\documentclass[12pt,aip,pop,reprint]{revtex4-2}

\usepackage{graphicx}
\usepackage{hyperref}
\hypersetup{colorlinks=true,linkcolor=blue,citecolor=blue}
\usepackage{bm}
\usepackage{multirow}
\usepackage{mathtools}

\begin{document}

\title{Collective Thomson scattering in non-equilibrium laser produced two-stream plasmas}

\author{K.~Sakai}
    \email[]{kentaro.sakai@eie.eng.osaka-u.ac.jp}
    \affiliation{Graduate School of Engineering, Osaka University, 2-1 Yamadaoka, Suita, Osaka 565-0871, Japan}
\author{S.~Isayama}
    \affiliation{Department of Physics, National Central University, No. 300, Jhongda Rd., Jhongli, Taoyuan 320, Taiwan}
    \affiliation{Department of Earth System Science and Technology, Kyushu University, 6-1 Kasuga-Koen, Kasuga, Fukuoka 816-8580, Japan}
\author{N.~Bolouki}
    \affiliation{Department of Physics, National Central University, No. 300, Jhongda Rd., Jhongli, Taoyuan 320, Taiwan}
    \affiliation{Center for Plasma and Thin Film Technologies, Ming Chi University of Technology, 84 Gungjuan Rd. Taishan Dist. New Taipei City 24301, Taiwan}
\author{M.S.~Habibi}
    \affiliation{Department of Physics, National Central University, No. 300, Jhongda Rd., Jhongli, Taoyuan 320, Taiwan}
\author{Y.L.~Liu}
    \affiliation{Department of Physics, National Central University, No. 300, Jhongda Rd., Jhongli, Taoyuan 320, Taiwan}
\author{Y.H.~Hsieh}
    \affiliation{Department of Physics, National Central University, No. 300, Jhongda Rd., Jhongli, Taoyuan 320, Taiwan}
\author{H.H.~Chu}
    \affiliation{Department of Physics, National Central University, No. 300, Jhongda Rd., Jhongli, Taoyuan 320, Taiwan}
    \affiliation{Center for High Energy and High Field Physics, National Central University, No. 300, Jhongda Rd., Jhongli, Taoyuan 320, Taiwan}
\author{J.~Wang}
    \affiliation{Department of Physics, National Central University, No. 300, Jhongda Rd., Jhongli, Taoyuan 320, Taiwan}
    \affiliation{Center for High Energy and High Field Physics, National Central University, No. 300, Jhongda Rd., Jhongli, Taoyuan 320, Taiwan}
    \affiliation{Institute of Atomic and Molecular Sciences, Academia Sinica, P. O. Box 23-166, Taipei, Taiwan}
\author{S.H.~Chen}
    \affiliation{Department of Physics, National Central University, No. 300, Jhongda Rd., Jhongli, Taoyuan 320, Taiwan}
\author{T.~Morita}
    \affiliation{Department of Advanced Energy Engineering Science, Kyushu University, 6-1 Kasuga-Koen, Kasuga, Fukuoka 816-8580, Japan}
\author{K.~Tomita}
    \affiliation{Department of Applied Science for Electronics and Materials, Kyushu University, 6-1 Kasuga-Koen, Kasuga, Fukuoka 816-8580, Japan}
\author{R.~Yamazaki}
    \affiliation{Department of Physics and Mathematics, Aoyama Gakuin  University, 5-10-1 Fuchinobe, Sagamihara, Kanagawa 252-5258, Japan}
    \affiliation{Institute of Laser Engineering, Osaka University, 2-6 Yamadaoka, Suita, Osaka 565-0871, Japan}
\author{Y.~Sakawa}
\affiliation{Institute of Laser Engineering, Osaka University, 2-6 Yamadaoka, Suita, Osaka 565-0871, Japan}
\author{S.~Matsukiyo}
    \affiliation{Department of Earth System Science and Technology, Kyushu University, 6-1 Kasuga-Koen, Kasuga, Fukuoka 816-8580, Japan}
\author{Y.~Kuramitsu}
    \email[]{kuramitsu@eei.eng.osaka-u.ac.jp}
    \affiliation{Graduate School of Engineering, Osaka University, 2-1 Yamadaoka, Suita, Osaka 565-0871, Japan}
    \affiliation{Institute of Laser Engineering, Osaka University, 2-6 Yamadaoka, Suita, Osaka 565-0871, Japan}

\date{\today}

\begin{abstract}
We investigate collective Thomson scattering (CTS) in two-stream non-equilibrium plasmas analytically, numerically and experimentally. In laboratory astrophysics, CTS is a unique tool to obtain local plasma diagnostics. While the standard CTS theory assumes plasmas to be linear, stationary, isotropic and equilibrium, it is often nonlinear, non-stationary, anisotropic, and non-equilibrium in high energy phenomena relevant to laboratory astrophysics. We theoretically calculate and numerically simulate the CTS spectra in two-stream plasmas as a typical example of non-equilibrium system in space and astrophysical plasmas. The simulation results show the feasibility to diagnose two-stream instability directly via CTS measurements. In order to confirm the non-equilibrium CTS analysis, we have been developing experimental system with high repetition rate table top laser for laboratory astrophysics. 
\end{abstract}

\maketitle

\section{Introduction}
\label{sec_intro}

Laboratory astrophysics is a research field, where space and astrophysical phenomena are reproduced experimentally in laboratories. In space plasmas, in-situ measurements with spacecrafts provides us the local and microscopic information of plasmas and electric/magnetic fields, however, it is hard to observe global structure of phenomena. In contrast, while imaging of astrophysical objects with telescope provides us the global and macroscopic information of phenomena, there is no local and microscopic information of plasmas since they are inaccessible. In laboratories, we can access both global and local information at the same time \cite{kuramitsu2012ppcf}. Besides this, while we have to image the emissions from the astrophysical phenomena, we can use external light sources and particle beams to diagnose the phenomena in laboratories. These are the significant advantages of laboratory astrophysics, and highly challenging in space and astrophysical plasmas \cite{kuramitsu2012ppcf,sakawa2016apx}. For instance, we have investigated collisionless shocks and magnetic reconnections relevant to space and astrophysical phenomena, such as supernova remnants, earth's bow shocks, solar flares, stellar winds, and aurorae, with Gekko XII (GXII) laser facility \cite{kuramitsu2011prl, kuramitsu2018ncom,morita2013pop,morita2019pop,bolouki2019hedp,sakawa2017hedp}, however, the number of shots is very limited due to the low repetition rate of GXII (a few shots per day) and there exist only several large facilities like GXII in the world. Therefore, the opportunities of experiments on laboratory astrophysics are limited. We are motivated to use high repetition-tabletop lasers since there are many more facilities to obtain much more data on laboratory astrophysics. We also extend the laboratory astrophysics with short pulse lasers. As the first step, we match the intensity of the tabletop lasers to that of high power lasers and confirm that the plasma with similar density and temperature can be obtained. We will study the plasma dynamics in the future. 

As mentioned above, in laboratories global and local information of phenomena can be obtained simultaneously. Collective Thomson scattering (CTS) is a unique tool to measure local plasma quantities of the density, velocity, and temperature both of electrons and ions \cite{morita2013pop, morita2019pop, bolouki2019hedp,sakawa2017hedp,ross2012pop}. A conventional CTS analysis assumes plasmas to be linear, stationary, equilibrium, and stable\cite{froula2011}, however, in many high energy space and astrophysical phenomena as well as laser-produced plasmas relevant to laboratory astrophysics, such as collisionless shocks and magnetic reconnections, are highly nonlinear, non-stationary, and non-equilibrium, and show various kinds of instabilities. Hence, the conventional CTS analysis may not be appropriate for such plasmas. For example, in the presence of a high-Mach number collisionless shock, some part of upstream plasma can be reflected at the shock front, and thus, in the upstream region of collisionless shock two-stream plasmas are often observed \cite{burgess2012ssr}. The two-stream plasmas can be unstable, and various wave activities resulting from the instabilities play essential roles on particle acceleration and generation of cosmic rays. We have analytically as well as numerically investigated the CTS in nonlinear, non-stationary, non-equilibrium, and unstable plasmas \cite{matsukiyo2016jpcs}. We further develop the CTS analysis for such non-equilibrium plasmas. So far, Thomson scattering from non-Maxwellian plasmas has been extensively investigated for super-Gaussian \cite{zheng1997pop,milder2019pop}, Spitzer-H\"{a}rm \cite{henchen2018prl,henchen2019pop}, and kappa distribution functions \cite{saito2000ag}. In this study, we consider two-plasma states as an example of such non-equilibrium plasmas and focus on CTS spectrum in the presence of the two-stream instability as well as the high energy components. The investigations of non-Maxwellian distribution functions \cite{zheng1997pop,milder2019pop,henchen2018prl,henchen2019pop,saito2000ag} in the past focus on the distribution functions symmetric about $v=0$. We consider here the two-stream plasmas that often seen in the upstream of collisionless shocks, where the distribution function is asymmetric. The two-stream plasmas and the resultant instabilities are significant since the relevant wave dynamics play essential roles in particle acceleration. One of our long term goals is the investigation of the origins of cosmic rays; we would like to understand the particle acceleration at collisionless shocks in a controlled manner in laboratories. To this end, in this paper we study the two-plasma states either with the different drift velocities or different temperatures. The latter can express a plasma with high energy component. In reality in space it should be mixture of these two and can be much more complicated. We start from the well-established CTS theory and extend it with two Maxwellian distributions as a typical example of upstream plasmas, which can be unstable or with high-energy component. Observing the instabilities and high energy component in collisionless shock via CTS will be an essential step toward understanding the particle acceleration in the universe.

When the velocity difference of two plasmas is not large, electron distribution functions overlap each other. The Landau damping at two peaks of CTS will be different from that in the presence of a single plasma, and the CTS spectra will change the form. Moreover, when the velocity difference of two plasmas is larger than thermal velocity of plasmas, two-stream instabilities grow \cite{francis1986pof}. It is considered that one of the peaks in CTS spectrum is enhanced when the two-stream instability takes place \cite{matsukiyo2016jpcs}. In order to understand the CTS from the non-equilibrium plasmas, we theoretically investigate the CTS spectra from two-stream plasmas in Sec. \ref{sec_theory}. In Sec. \ref{sec_simulation}, we numerically investigate the CTS in the presence of two-stream instability. In Sec. \ref{sec_experiment}, we introduce our experimental approach to verify the non-equilibrium CTS in laser produced plasmas. To verify CTS in non-equilibrium plasma, large laser facilities are not convenient due to the low repetition rate of laser. In Sec. \ref{sec_summary} we summarize our research. 

\section{Theoretical spectrum from two-stream plasmas}
\label{sec_theory}

\begin{figure*}
    \centering
    \includegraphics[clip,width=\hsize]{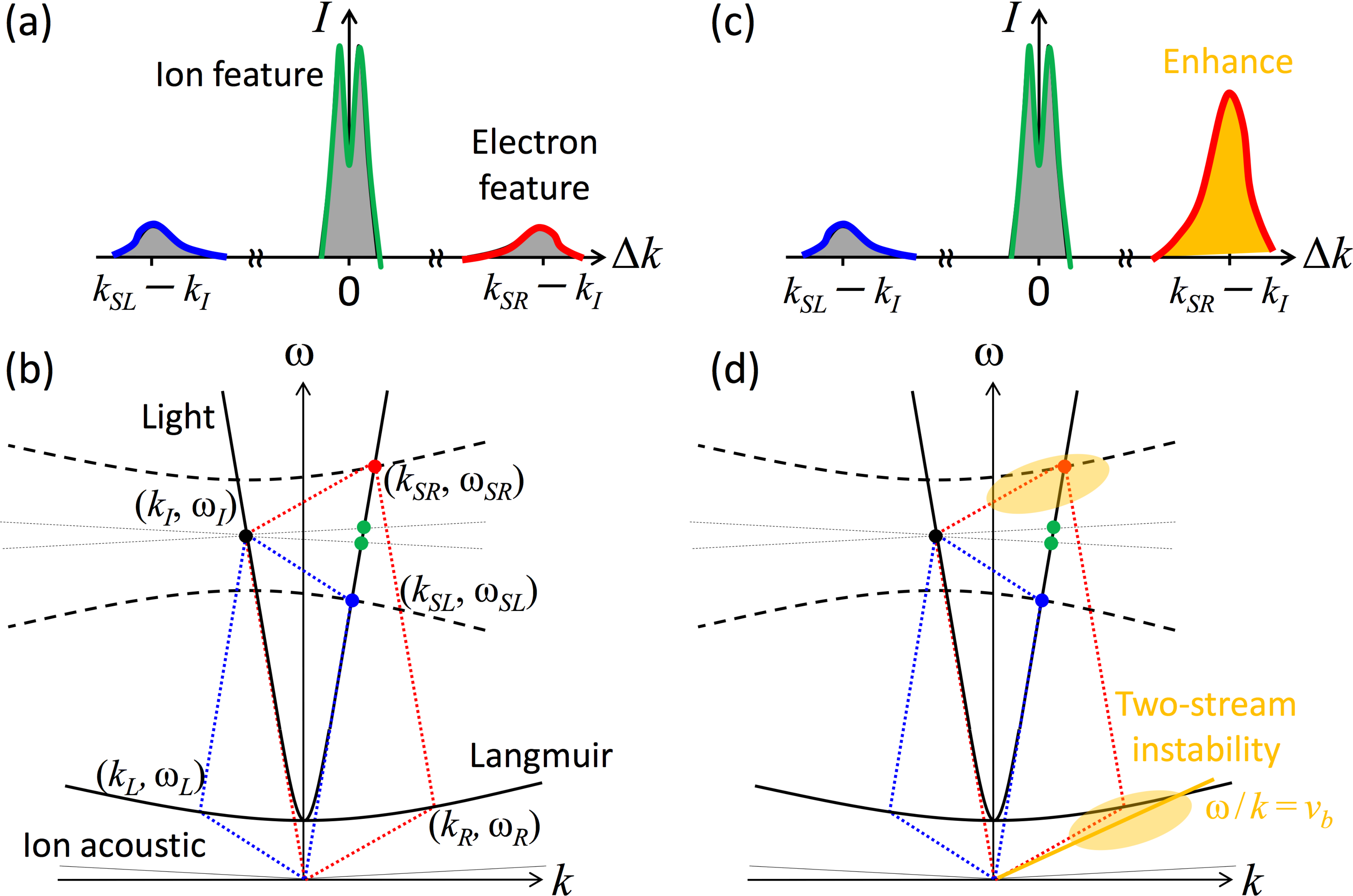}
    \caption{(a) Schematic image of CTS spectra with a single plasma. (b) Dispersion relations of light, Langmuir, and ion acoustic waves in a single plasma. (c) Schematic image of CTS spectra in two-stream plasmas. (d) Dispersion relations in two-stream plasmas. Two-stream instability grows in the orange region. When the region corresponds to one of the peaks of CTS, the peak is enhanced as shown in (c).}
    \label{fig_1}
\end{figure*}

Figures~\ref{fig_1}~(a) and (b) shows the schematic images of CTS spectra and the dispersion relation in the presence of a single plasma. An incident light wave with the frequency $\omega_I$ and the wavenumber $k_I$ can be parametrically scattered by the Langmuir waves and by the ion acoustic waves. As shown in Fig.~\ref{fig_1}~(a), while the light scattered by the ion acoustic waves corresponding to the CTS ion feature has the higher peak intensity ($I$) and the narrower spectral width, the light scatted by the Langmuir waves corresponding to the CTS electron feature has the lower and broader spectra, where the horizontal axis $\Delta k \equiv k_S-k_I$ shows the wavenumber difference between the scattered and incident waves and the subscripts $L$ and $R$ represent left and right. In the collective regime, there are two peaks in each feature coming from the scattering by the waves propagating in the same and opposite directions. For instance, the left  peak of the electron feature in Fig.~\ref{fig_1}~(a) comes from the resonant interaction between the incident light, the scattered light $(k_{SL}, \omega_{SL})$, and the Langmuir wave $(k_{L}, \omega_{L})$ forming a parallelogram in Fig.~\ref{fig_1}~(b), where the incident and the Langmuir mode co-propagate. Similarly, the right peak of the electron feature comes from the other parallelogram composed by the incident $(k_{I}, \omega_{I})$, scatter $(k_{SR}, \omega_{SR})$, and Langmuir $(k_{R}, \omega_{R})$ waves in Fig.~\ref{fig_1}~(b), where the incident and the Langmuir mode counter-propagate. Figures~\ref{fig_1}~(c) and (d) show the same as Figs.~\ref{fig_1}~(a) and (b) except with two-stream instability. The velocity of beam component of plasma is expressed by the oblique line with the velocity $v_{b}$ in Fig.~\ref{fig_1}~(d). When the line intersects with the Langmuir branch, the two-stream instability can grow. Since the electron feature of CTS is the resonance interaction between the incident electromagnetic, electron plasma (Langmuir), and scattered electromagnetic waves, the amplitude of scatter wave or the peak intensity of CTS is proportional to the density fluctuation of Langmuir waves. The Langmuir waves enhanced by the two-stream instability will enhance the scattered wave amplitude. When the phase velocity of plasma wave observed in CTS is in unstable region in Fig.~\ref{fig_1}~(d) that is expressed as $\omega / k \sim v_{b}$, the plasma wave grows and the corresponding peak of CTS can be enhanced as shown in Fig.~\ref{fig_1}~(c). Since the electron distribution function becomes non-Maxwellian with a beam component of plasma, the shape of electron distribution function also changes the CTS spectra. The two-stream instability is not included in the theory, but we simply include two plasmas in the theory. In this section, we discuss CTS spectrum with electron distribution functions different from Maxwellian. We consider two cases where two plasmas coexist either with finite relative velocity or with finite temperature difference. 

We calculate the scattering form factor assuming two-stream plasmas. The spectrum shape of Thomson scattering is related to the scattering form factor, which is expressed as 
\begin{equation}
S(k,\omega) = \frac{2\pi}{k} \left[\left|1-\frac{\chi_{e}}{\epsilon}\right|^{2} f_{e} \left(\frac{\omega }{k}\right) + Z \left|\frac{\chi_{e}}{\epsilon}\right|^{2} f_{i} \left(\frac{\omega}{k}\right)\right],
\label{eq_form}
\end{equation}
where $k$, $\omega$, $\chi_{e}$, $\epsilon$, $Z$, $f_{e}(v)$, and $f_{i}(v)$ are the scattering wavenumber, scattering frequency, electron susceptibility, permittivity, ion valence, electron distribution function, and ion distribution function, respectively \cite{froula2011}. This formula assumes quasi-equilibrium plasma. In this paper, we consider only the electron feature of CTS, and ignore the second term of the right hand side. We assume electron distribution functions as superposition of two Maxwellian distributions, which is written as 
\begin{equation}
f_{e 1+2}(v) = \sum_{j} \frac{n_{ej}}{n_{e}} f_{ej}(v),
\label{eq_fe}
\end{equation}
where $n_{ej}$, $f_{ej}(v)$, and $n_{e}$ are the electron density and electron distribution function of the $j$-th plasma species, and total electron density, respectively. The electron distribution function of the $j$-th plasma is given by $f_{ej}(v) = \sqrt{1/(\pi v_{tej}^{2})} \exp (-(v-v_{dj})^{2}/v_{tej}^{2})$, where $v_{tej}$ and $v_{dj}$ are $j$-th electron thermal velocity and $j$-th drift velocity, respectively. The $j$-th electron thermal velocity is written as $v_{tej} = \sqrt{2 k T_{ej} / m_{e}}$, where $T_{ej}$ is $j$-th electron temperature. With the electron distribution function, the electron susceptibility is given by 
\begin{equation}
\begin{split}
\chi_{e} & = \frac{4 \pi e^{2} n_{e}}{m_{e} k^{2}} \int_{-\infty}^{\infty} \frac{\frac{\partial f_{e 1+2}}{\partial v}}{\frac{\omega}{k} - v} dv \\
& = -\frac{4 \pi e^{2} }{ m_{e} k^{2}} \sum_{j} \left[ \frac{n_{ej}}{v_{tej}^{2}} Z'\left(\frac{\frac{\omega}{k} - v_{dj}}{v_{tej}}\right) \right ],
\label{eq_chie}
\end{split}
\end{equation}
where $e$, $m_{e}$, and $Z'(\xi)$ are the elementary charge, electron mass, and the derivative of the plasma dispersion function, respectively. Since the integrand has a singular point at $v=\omega /k$ on the integration path, the imaginary part is the residue at the singular point, and the real part is the integral value outside the singular point. As Eq.~(\ref{eq_chie}) shows, the imaginary part of the electron susceptibility is proportional to the derivative of electron distribution function and this value represents Landau damping \cite{froula2011}. When the relative drift velocity between two plasmas is much larger than the thermal velocities ($\Delta v = |v_{d1} - v_{d2}| \lesssim v_{tej}$), the scattering form factor is not appropriate to express the CTS spectrum due to the quasi-equilibrium assumption. Nevertheless, we also calculate the case with $\Delta v \gg v_{tej}$ for the comparison purpose. 

\begin{figure*}
    \centering
    \includegraphics[clip,width=\hsize]{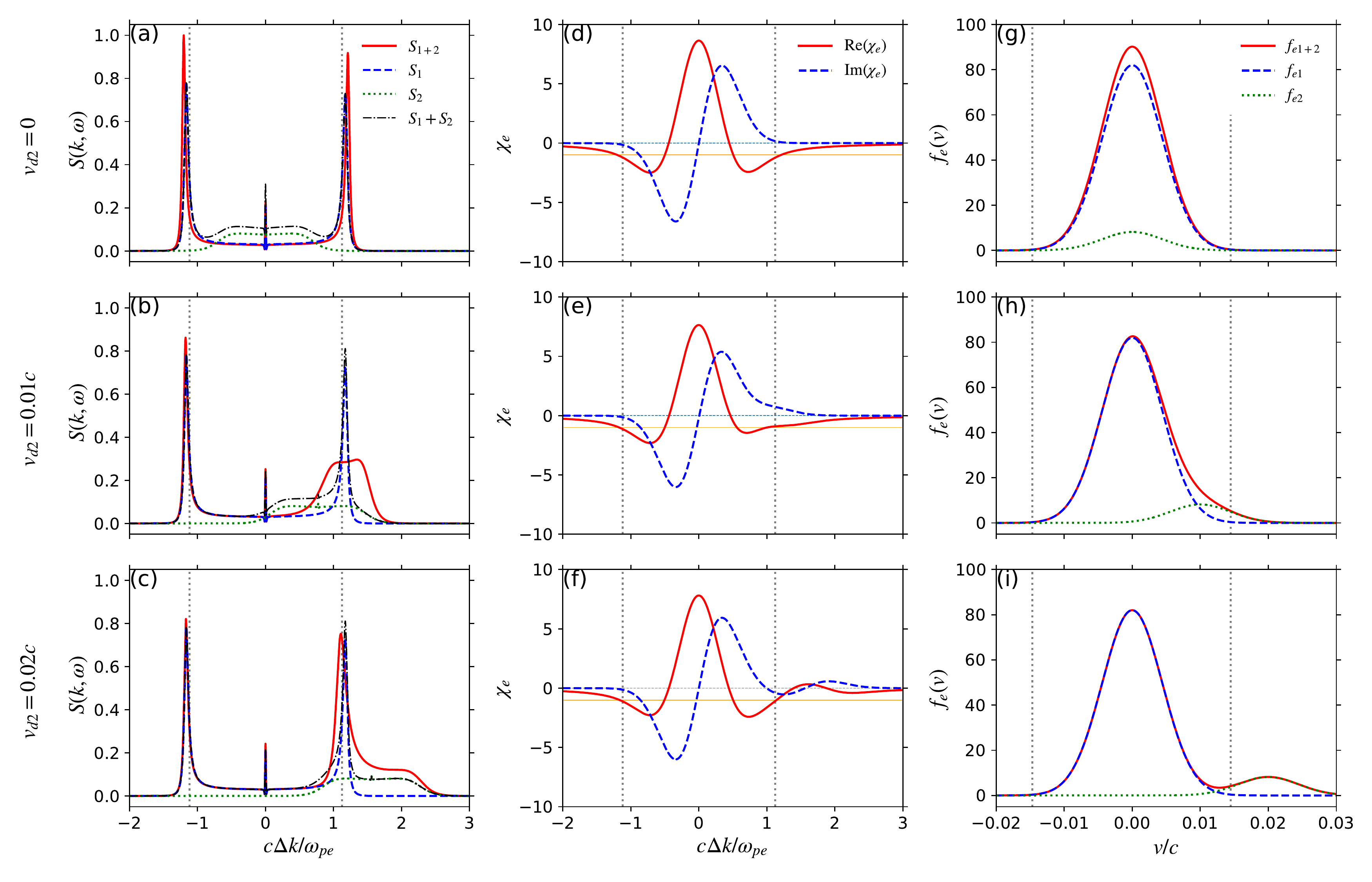}
    \caption{(a)-(c) Theoretical CTS spectra from two-stream plasmas. (d)-(f) Electron susceptibilities. (g)-(i) Electron distribution functions. (a), (d), (g) $v_{d2} = 0$, (b), (e), (h) $v_{d2} = 0.01 c$ $(\Delta v / v_{te1} = 1.6)$, (c), (f), (i) $v_{d2} = 0.02 c$ $(\Delta v / v_{te1} = 3.2)$. The dotted vertical lines in (a)-(f) and (g)-(i) represent the wavenumbers and phase velocities of the CTS peaks with the $j=1$ plasma, respectively.}
    \label{fig_2}
\end{figure*}

First, we consider the case with finite relative drift velocity. We plot Eqs.~(\ref{eq_form}), (\ref{eq_fe}), and (\ref{eq_chie}) by changing the relative drift velocity between two plasmas keeping the other parameters same in Fig.~\ref{fig_2}. The solid, dashed, and dotted curves in Figs.~\ref{fig_2}~(a)-(c) represent the scattering form factors from two plasmas considering the overlap of electron distribution functions ($S_{1+2}$), from the $j=1$ plasma ($S_{1}$) and from the $j=2$ plasma ($S_{2}$), respectively. The solid and dashed curves in Figs.~\ref{fig_2}~(d)-(f) represent the real and imaginary parts of the electron susceptibility, respectively. The horizontal axes in Figs.~\ref{fig_2}~(a)-(f) are the difference between scattered and incident wavenumber, $\Delta k = k_{S} - k_{I}$. The solid, dashed, and dotted curves in Figs.~\ref{fig_2}~(g)-(i) represent the electron distribution functions, $f_{e 1+2}$, $f_{e1}$, and $f_{e2}$, respectively, which is normalized as $\int_{-\infty}^{\infty} f_{e 1+2}(v/c) dv = 1$. We set the parameter of plasmas as $T_{e1}=10~\mathrm{eV}$, $T_{e2}=10~\mathrm{eV}$, $n_{e1}=1\times10^{18}~\mathrm{cm^{-3}}$, $n_{e2}=1\times10^{17}~\mathrm{cm^{-3}}$, $v_{d1}=0$, $\theta = 80~\mathrm{degrees}$, where $\theta$ is the scattering angle. We change $v_{d2}$ as $0$, $0.01c$, and $0.02c$, where $c$ is speed of light. The scattering $\alpha$ is given by $1/(k\lambda_{D})$, where $\lambda_{D}$ is the Debye length. When $\alpha \gtrsim 1$, Thomson scattering from collective plasma waves is dominant \cite{froula2011}. When $\alpha \ll 1$, Thomson scattering is non-collective and comes from the random electron motions \cite{froula2011}. Collective and non-collective scatterings show two peaks and a single peak in their spectra, respectively. As shown in Fig.~\ref{fig_1}~(b), the peak scattered frequency $\omega_{S}$ and wavenumber $k_{S}$ are determined from the intersections between the  light waves $\left( \omega^{2} = \omega_{pe}^{2} + c^{2}k^{2} \right)$ and shifted Langmuir waves $\left[(\omega - \omega_{I})^{2} = \omega_{pe}^{2} + v_{te}^{2}(k-k_{I})^{2} \right]$, where $\omega_{pe}$, $\omega_{I}$, $k_{I}$, and $v_{te}$ are the electron plasma frequency, incident frequency, incident wavenumber, and $\sqrt{3/2} v_{te1}$, respectively. The scattered wavenumber is calculated from the dispersion relations with the assumption of $\omega \sim \pm ck$, 
\begin{equation}
\begin{split}
k_{S} \sim & \frac{c^{2} - v_{te}^{2} \cos \theta}{c^{2} - v_{te}^{2}} k_{I} \\
& \pm \sqrt{\frac{4 v_{te}^{2} \sin^{2} (\theta/2) \left[ c^{2} - v_{te}^{2} \cos^{2} (\theta/2) \right] }{ \left( c^{2} - v_{te}^{2} \right)^{2}} k_{I}^{2} + \frac{\omega_{pe}^{2}}{c^{2} - v_{te}^{2}}}.
\label{eq_k}
\end{split}
\end{equation} 
The dotted vertical lines in Figs.~\ref{fig_2}~(a)-(f) represent the peak scattered wavenumber of $j=1$ plasma. The peak phase velocities of Langmuir wave are written as 
\begin{equation}
v_{pj} = \frac{\omega_{S}-\omega_{I}}{|{\bf k}_{\mathrm{S}}-{\bf k}_{\mathrm{I}}|} = \frac{\sqrt{\omega_{pe}^{2} + c^2 k_{S}^2} - \sqrt{\omega_{pe}^{2} + c^2 k_{I}^2}}{\sqrt{k_{I}^{2} + k_{S}^{2} -2 k_{I} k_{S} \cos \theta}}. 
\label{eq_vphi}
\end{equation}
The dotted vertical lines in Figs.~\ref{fig_2}~(g)-(i) represent the phase velocity of the corresponding peaks of CTS spectra (Figs.~\ref{fig_2}~(a)-(c)) in the $j=1$ plasma. The exact peaks of Eq.~(\ref{eq_form}) are determined by the absolute value term, $|1-(\chi_{e}/\epsilon)|^{2}$. The permittivity is approximated to $1 + \chi_{e}$ at the peak wavenumber of the electron feature, and then, the absolute value term becomes $1/|1 + \chi_{e}|^{2} = 1/\left((1 + \mathrm{Re}(\chi_{e}))^{2} + \mathrm{Im}(\chi_{e})^{2}\right)$. The peaks are at the wavenumber with the smallest denominator of the absolute value term, i.e.~$\mathrm{Re}(\chi_{e}) \sim -1$ and $\mathrm{Im}(\chi_{e}) \sim 0$. The horizontal solid and dashed lines in Figs.~\ref{fig_2}~(d)-(f) represent $\chi_{e} = -1$ and $0$, respectively. In the presence of a single plasma, this corresponds to the resonant condition, $\epsilon = 0$. 

Figures~\ref{fig_2}~(a), (d), and (g) show the results without the relative drift, i.e.~two peaks are nearly symmetric. As shown in Fig.~\ref{fig_1}~(b), the phase velocity of Langmuir wave in right peak ($\omega_{R}/k_{R}$) is slightly different from that in left peak ($\omega_{L}/k_{L}$). This results in the difference of Landau damping between two peaks.The wavenumbers where $\mathrm{Re}(\chi_{e}) \sim -1$ and $\mathrm{Im}(\chi_{e}) \sim 0$ slightly deviate from the vertical dotted lines since the total electron density is $1.1$ times larger than that of $j=1$ plasma. Figures~\ref{fig_2}~(b), (e) and (h) are the cases when $0<v_{d2}<v_{p1}$. Derivative of electron distribution function at the peak phase velocities in $v>0$ in Fig.~\ref{fig_2}~(h) are steeper than that in Fig.~\ref{fig_2}~(g), and the peak intensity of solid curve for $\Delta k > 0$ in Fig.~\ref{fig_2}~(b) is lower than that of dashed one. The right peak is attenuated by Landau damping when $0<v_{d2}<v_{p1}$. Figure~\ref{fig_2}~(e) shows $\mathrm{Re}(\chi_{e}) \sim -1$ and $\mathrm{Im}(\chi_{e}) > 0$ around the dotted vertical line for $\Delta k > 0$, resulting in the peak broadening and attenuation. As the right peak intensity is low despite the larger electron distribution function at the peak phase velocity, the effect of Landau damping is quite larger than that in Fig.~\ref{fig_2}~(g). Figures~\ref{fig_2}~(c), (f), and (i) show the case when $v_{d2}>v_{p1}$ and $v_{d2} \gtrsim v_{te1}$. The derivative of electron distribution function at the peak phase velocities in $v>0$ is positive in Fig.~\ref{fig_2}~(i) (opposite to Figs.~\ref{fig_2}~(g) and (h)). Thus, the right peak can be enhanced due to Landau resonance, however, there is no such amplification comparing Fig.~\ref{fig_2}~(a). The wave growth is not included in the conventional CTS theory. Note that the last case ($v_{d2} = 0.02 c$) is unstable and will change the electron distribution function in a short time. In order to consider the nonlinear evolutions of the two-stream instability, numerical simulations are necessary. 

\begin{figure*}
    \centering
    \includegraphics[clip,width=\hsize]{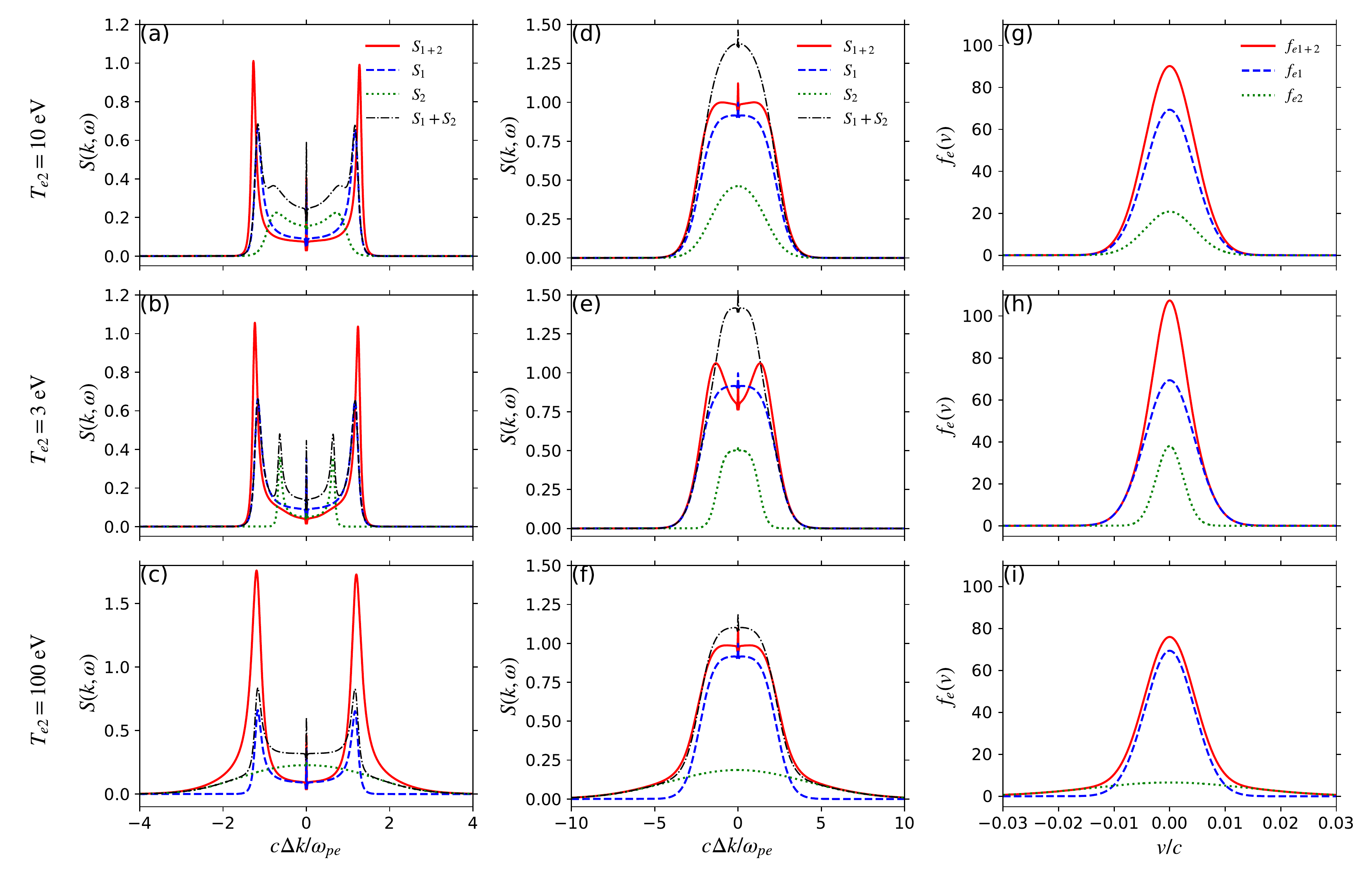}
    \caption{(a)-(f) Theoretical CTS spectra from two-stream plasmas changing the temperature. (g)-(i) Electron distribution functions. (a)-(c) Collective scattering ($n_{e1} = 1 \times 10^{17} ~ \mathrm{cm^{-3}}$), (d)-(f) non-collective scattering ($n_{e1} = 1 \times 10^{16} ~ \mathrm{cm^{-3}}$). (a), (d), (g) $T_{e2} = 10~\mathrm{eV}$, (b), (e), (h) $T_{e2} = 3~\mathrm{eV}$, (c), (f), (i) $T_{e2} = 100~\mathrm{eV}$.}
    \label{fig_3}
\end{figure*}

Comparing the solid, dashed, and dotted curves in Figs.~\ref{fig_2}~(a)-(c) and Figs.~\ref{fig_3}~(a)-(f), $S_{1+2}$ (solid ones) are different from simple sum of $S_{1}$ and $S_{2}$ (dashed and dotted ones). The difference between $S_{1+2}$ and $S_{1} + S_{2}$ is the electron susceptibility in Eq.~(\ref{eq_chie}). The electron susceptibility is proportional to the derivative of electron distribution function, which is the factor to determine Landau damping. Thus, the Landau damping is a major factor to determine the spectrum shape of $S_{1+2}$ and $S_{1} + S_{2}$ in Fig.~\ref{fig_2}. When $v_{d2} < v_{te1}$, the effect of Landau damping seems to be appropriate. However, when there is a positive derivative in the electron distribution function, wave growth due to Landau resonance is not taken into account; the peak is damped in Fig.~\ref{fig_2}~(c). Moreover, as shown in Appendix \ref{sec_appendix}, the scattering form factor has nothing to do with the instability but is mathematically determined. Although the scattering form factor assumes quasi-equilibrium plasma, the electron distribution functions shown in Fig.~\ref{fig_2}~(i) is rather unstable. Such electron distribution functions change in a short time and the scattering form factor is not appropriate for such plasmas. In such cases, we need numerical simulations. 

The sign of derivative of distribution function, which determines wave damping and growth, is not considered in the theory since the squared absolute value of susceptibility and permittivity is calculated. The conventional theory only describes the Landau damping. The squared absolute value comes from the ensemble average of electron density, which defines the scattering form factor and is given by
\begin{equation}
    S({\bf k}, \omega) \equiv \lim_{V\rightarrow \infty,T\rightarrow \infty} \frac{1}{VT} \left \langle \frac{n_e({\bf k}, \omega), n_e^*({\bf k}, \omega)}{n_{e0}} \right \rangle ,
\end{equation}
where $n_{e0}$, $V$, and $T$ represent the mean electron density, scattering volume, and scattering time, respectively \cite{froula2011}. In order to construct completely non-equilibrium theory, we must start with the most fundamental equation or the Vlasov equation.

Now we consider the electron distribution functions with the same drift velocity but with different temperatures, which is also the case of the high energy components of distribution function with $T_{e1} < T_{e2}$ and our experiment shown later with $T_{e1} > T_{e2}$. We plot Eqs.~(\ref{eq_form}) and (\ref{eq_fe}) by changing $T_{e2}$ keeping the other parameters the same as in Fig.~\ref{fig_3}. We set the parameter of plasmas as $T_{e1}=10~\mathrm{eV}$, $v_{d1}=v_{d2}=0$, $\theta = 30~\mathrm{degrees}$. Figures~\ref{fig_3}~(a)-(c) and Figs.~\ref{fig_3}~(d)-(f) show the cases of collective scattering ($n_{e1} = 1 \times 10^{17} ~ \mathrm{cm^{-3}}$ and $\alpha_{1} = 2.2$) and rather non-collective scattering ($n_{e1} = 1 \times 10^{16} ~ \mathrm{cm^{-3}}$ and $\alpha_{1} = 0.7$), respectively. The rather non-collective scattering is shown here because the scattering from the high energy component or non-thermal component in an upstream plasma of collisionless shock can be non-collective. The ratio of $n_{e2}$ to $n_{e1}$ is 0.3. We change $T_{e2}$ as 10 for reference, 3, and 100 eV. Figures~\ref{fig_3}~(g)-(i) show the normalized electron distribution functions common to both cases. 
First, we consider collective scattering. In Fig.~\ref{fig_3}~(a), $T_{e1} = T_{e2}$ and $\alpha_{2} = 1.2$, and thus, the scattering from $j=2$ plasma is also collective. Three curves in Fig.~\ref{fig_3}~(a) show two peaks. In Fig.~\ref{fig_3}~(b), $T_{e1} > T_{e2}$ and $\alpha_{2} = 2.2$. Although the electron distribution function $f_{e 1+2}$ in Fig.~\ref{fig_3}~(h) has more low energy electrons than that in Fig.~\ref{fig_3}~(g), the spectra are similar to that in Fig.~\ref{fig_3}~(a). Figure~\ref{fig_3}~(c) shows the case when $T_{e1} < T_{e2}$ and  $\alpha_{2} = 0.38$. As $f_{e 1+2}$ in Fig.~\ref{fig_3}~(i) has more high energy component than that in Fig.~\ref{fig_3}~(g), $S_{1+2}$ in Fig.~\ref{fig_3}~(c) is asymptotic to $S_{2}$ when $|\Delta k| \gg 0$. Now, we consider non-collective scattering. In Fig.~\ref{fig_3}~(d), while $S_{1}$ and $S_{2}$ show a single peak, $S_{1+2}$ shows two peaks. As $\alpha_{2} = 0.38$ in Fig.~\ref{fig_3}~(d), $S_{2}$ is also non-collective. Figure~\ref{fig_3}~(e) shows the case when $T_{e1} > T_{e2}$ and $\alpha_{2} = 0.7$. The two peaks of $S_{1+2}$ in Fig.~\ref{fig_3}~(e) are more prominent than that in Fig.~\ref{fig_3}~(d). Figure~\ref{fig_3}~(f) shows the case when $T_{e1} < T_{e2}$ and $\alpha_{2} = 0.12$, and $S_{1+2}$ is asymptotic to $S_{2}$ as seen in Fig.~\ref{fig_3}~(c). 

When the both plasmas are at rest but with $T_{e1} > T_{e2}$ in collective scattering, since the derivatives of $f_{e 1+2}$ in Fig.~\ref{fig_3}~(h) at the resonant velocities are not greatly different from that in Figs.~\ref{fig_3}~(g), the second plasma has rather minor effect, as seen in Fig.~\ref{fig_3}~(a) and (b). In Fig.~\ref{fig_3}~(d), $S_{1+2}$ shows two peaks, which are the characteristics of collective scattering, although $S_{1}$ and $S_{2}$ are rather non-collective. Since the total electron density is larger than that of $S_{1}$ and $S_{2}$, the $\alpha$ is higher and the scattering seems rather collective. When $T_{e1} > T_{e2}$, the electron distribution function $f_{e 1+2}$ in Fig.~\ref{fig_3}~(h) has the lower effective temperature and the larger effective $\alpha$ than that in Fig.~\ref{fig_3}~(g), resulting in more clear two peaks in Fig.~\ref{fig_3}~(e) than that in Fig.~\ref{fig_3}~(d). Regardless of the scattering process being collective or non-collective, $S_{1+2}$ asymptotically approaches to $S_{2}$ when $T_{e1} \ll T_{e2}$ as shown in Figs.~\ref{fig_3}~(c) and (f). 

\section{Numerical simulations}
\label{sec_simulation}

\begin{figure}
    \centering
    \includegraphics[clip,width=\hsize]{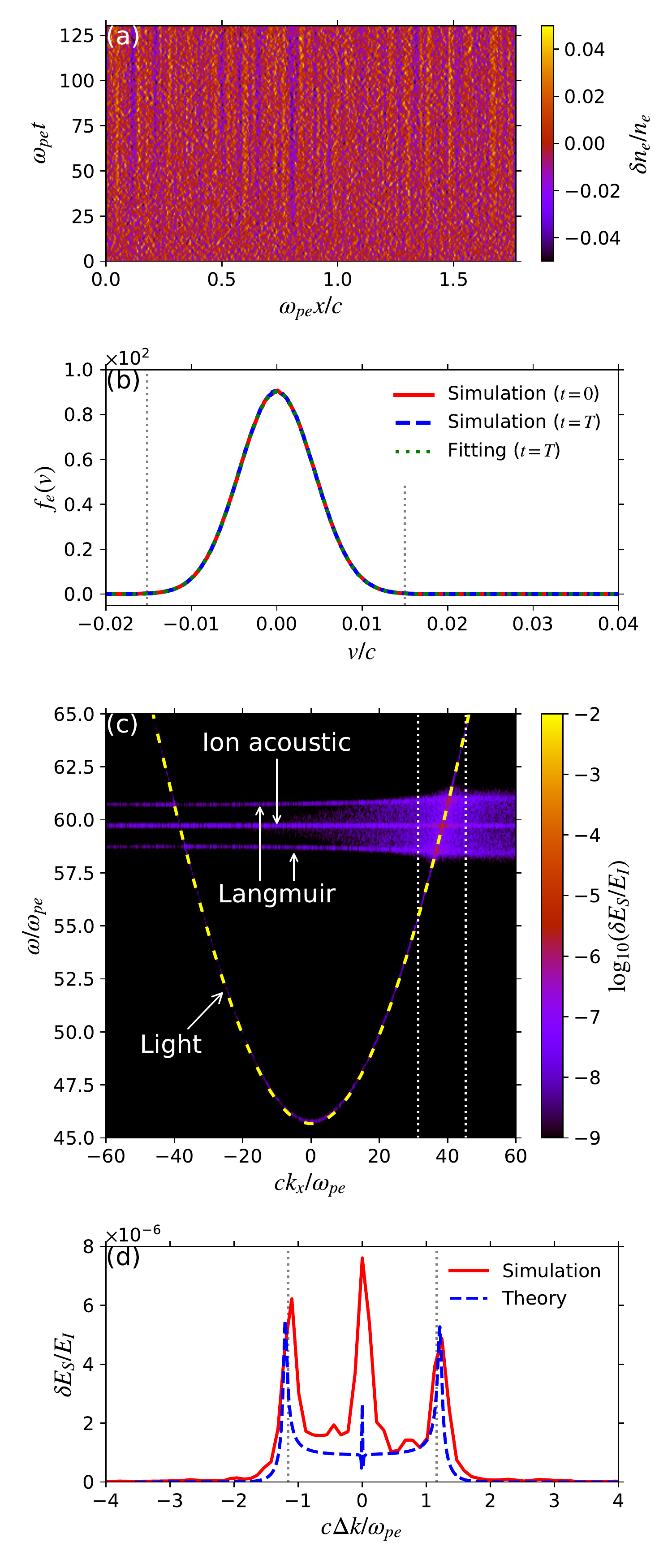} 
    \caption{Numerical results when $v_{d2}=0$. (a) Spatio-temporal evolution of the electron density fluctuation from the PIC simulation. (b) Distribution functions from the PIC simulation. (c) Dispersion relation from the FDTD simulation. (d) CTS spectra from the simulation and theory (Eq.~(\ref{eq_form})).}
    \label{fig_4}
\end{figure}

\begin{figure}
    \centering
    \includegraphics[clip,width=\hsize]{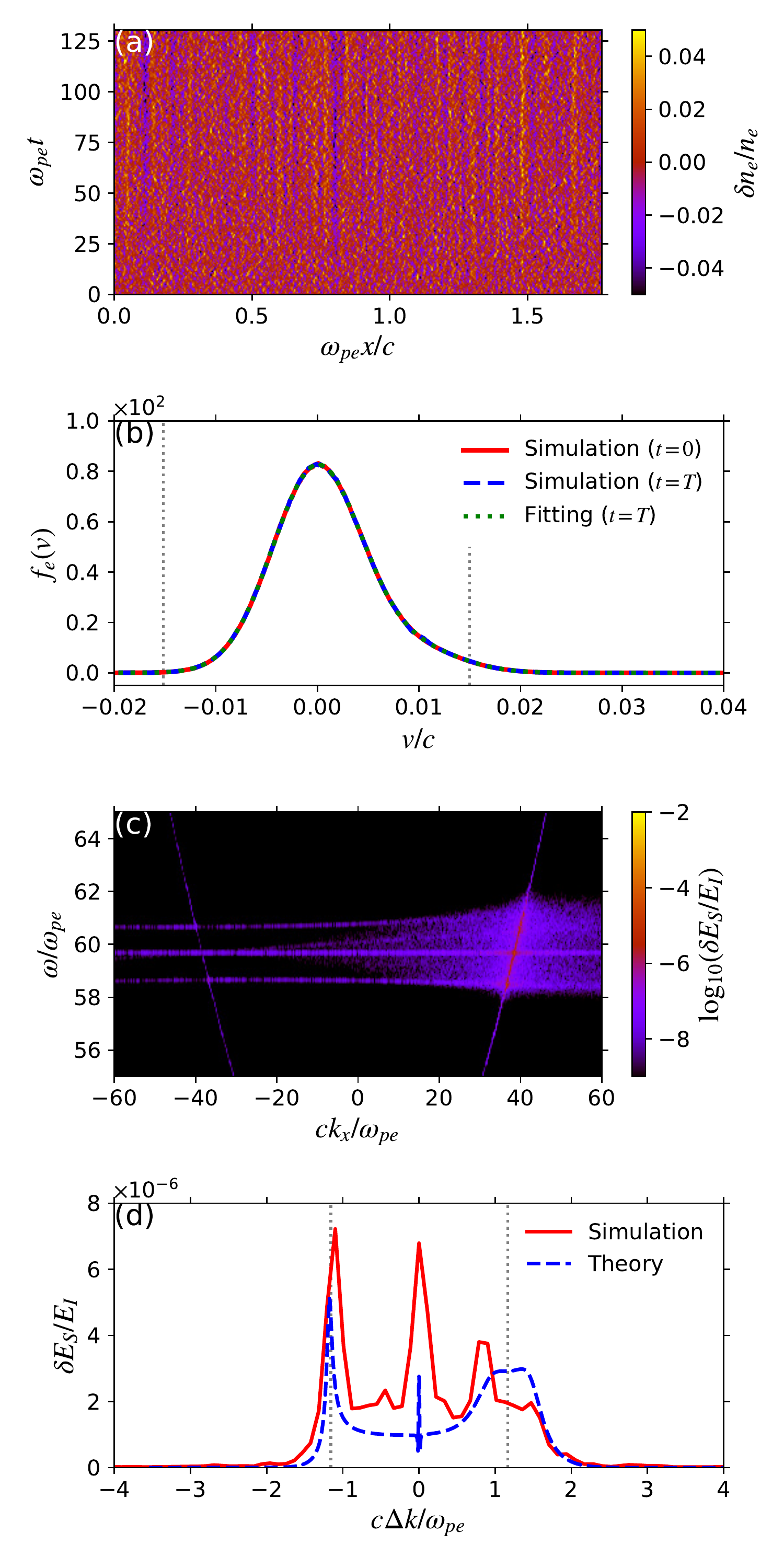} 
    \caption{Numerical results when $v_{d2}=0.01c$ $(\Delta v / v_{te1} = 1.6)$. (a) Electron density fluctuation. (b) Distribution functions. (c) Dispersion relation. (d) CTS spectra.} 
    \label{fig_5}
\end{figure}

\begin{figure}
    \centering
    \includegraphics[clip,width=\hsize]{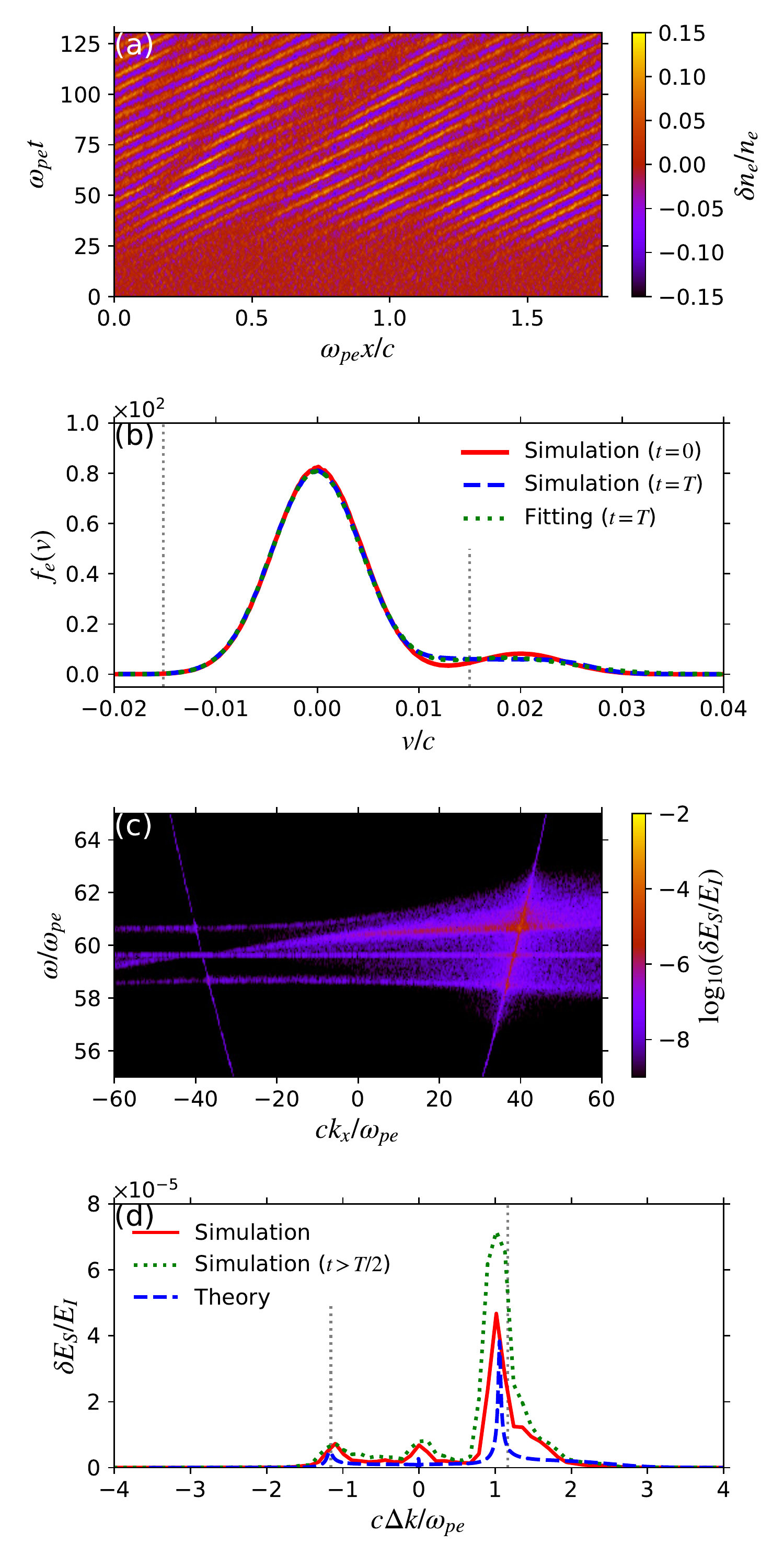} 
    \caption{Numerical results when $v_{d2}=0.02c$ $(\Delta v / v_{te1} = 3.2)$. (a) Electron density fluctuation. (b) Distribution functions. (c) Dispersion relation. (d) CTS spectra.} 
    \label{fig_6}
\end{figure}

Although we considered two-stream plasmas in the previous section, the conventional CTS theory assumes a plasma not far from the equilibrium. We calculate the CTS spectrum directly by solving wave equation. In the previous study\cite{matsukiyo2016jpcs}, the wave equation of the scattered waves is numerically solved by giving wave spectra of the incident electromagnetic wave and longitudinal fluctuations of a plasma, which are assumed before and after the development of the two-stream instability. In this study, we numerically calculate the density fluctuations in two-stream plasmas using particle-in-cell (PIC) simulations in order to express the growth of the two-stream instability. Using the density fluctuations obtained from the PIC simulations, we solve the wave equation of the scattered waves by finite-difference time-domain (FDTD) simulations. 

We numerically calculate two-stream interactions and obtain temporal and spatial evolutions of electron density fluctuations. We performed PIC simulations with EPOCH open source code \cite{arber2015ppcf}. We simulate the theoretical  configurations considered in the Sec. \ref{sec_theory}, where two Maxwellian plasmas exist: one is stationary high density plasma and the other is moving low density plasma. We calculate the time evolution of these plasmas. The parameters used in the PIC simulations are $n_{e1}=1 \times 10^{18}~\mathrm{cm^{-3}}$, $n_{e2}=1 \times 10^{17}~\mathrm{cm^{-3}}$, $T_{e1}=10~\mathrm{eV}$, $T_{e2}=10~\mathrm{eV}$, and $v_{d1} = 0$. We change $v_{d2}$ as $0$, $0.01c$, and $0.02c$. The grid size is $\Delta x =\lambda_{D}$ and the time step is $\Delta t = 0.45 \Delta x /c$. The number of grids is $N_{x} = 8192$ and the total number of time steps is $N_{t} = 65536$, corresponding to the system size $L = N_{x} \Delta x \sim 36c/\omega_{pe}$ and the computation time $T = N_{t} \Delta t \sim 130\omega_{pe}$. The number of particles in cell is 1000. 

We performed FDTD simulation with the electron density fluctuation, $\delta n_{e}$, calculated from the PIC simulation. The wave equation is written as \cite{matsukiyo2016jpcs} 
\begin{equation}
    \left(- c^{2} \frac{\partial^{2} }{ \partial x^{2} }+ \frac{\partial^{2} }{ \partial t^{2}} + \omega_{pe} \right) \delta \mathrm{{{\bf E}}_{S}} = 4 \pi e \frac{\partial }{ \partial t } \left( {\bf v}_{\mathrm{Ie}} \delta n_{e} \right),
\end{equation}
where $\delta {{\bf E}}_{\mathrm{S}}$ and $\mathrm{{\bf {v}}_{\mathrm{Ie}}}$ are the electric field of scattered wave and electron velocity determined by the electric field of incident wave \cite{matsukiyo2016jpcs}. The numerical parameters are the same as ones in the PIC simulations. The wavelength of incident wave is 532~nm and scattering angle is 80~degrees.

Figure~\ref{fig_4} represents the results of simulations when $v_{d2} = 0$ and these show the single plasma. Figure~\ref{fig_4}~(a) shows the spatio-temporal evolution of the electron density fluctuation during the entire computation time, and there is almost no drastic change of electron density over time. The vertical stripes show the ion acoustic waves. The solid and dashed curves in Fig.~\ref{fig_4}~(b) are the electron distribution functions from the PIC simulations when $t=0$ and $t=T$, respectively. We fit the dashed curve with two Maxwellian distributions, and the result is the dotted curve in Fig.~\ref{fig_4}~(b). The fitted curves are in good agreement with the simulated distribution functions at $t=T$. The dotted vertical lines are the peak phase velocities from Eq.~(\ref{eq_vphi}). Without the relative drift, there is no difference in electron distribution functions during the simulation. We numerically calculate the spatio-temporal evolution of $\delta {{\bf E}}_{\mathrm{S}}$ and perform the Fourier transform in all the $x-t$ space. Figure~\ref{fig_4}~(c) shows the amplitude of $\delta {{\bf E}}_{\mathrm{S}}$ in $\omega-k_{x}$ space, where $k_{x}$ is the incident wavenumber component parallel to the scattering vector, ${\bf k}$. The dashed curve in Fig.~\ref{fig_4}~(c) is the dispersion relation of electromagnetic waves or light waves in plasma. The peak at $(c k_{x} / \omega_{pe},~ \omega / \omega_{pe}) \sim (-38,~59)$ corresponds to the incident wave. The line extending right and left at $\omega \sim 59 \omega_{pe}$ is the dispersion relation of ion acoustic wave, and the curves at $\omega \sim (59 \pm 1) \omega_{pe}$ are the dispersion relations of Langmuir wave. Picking up $\delta E_{\mathrm{S}}$ on the dispersion relation of light waves, the CTS spectrum in terms of the wavenumber is obtained. The solid and dashed curves in Fig.~\ref{fig_4}~(d) are the CTS spectra obtained by numerical simulation and by the square root value of theoretical function, Eq.~(\ref{eq_form}), which is proportional to the scattered electric field, respectively. We obtain the simulated spectrum using all the computation time. The electron distribution function to calculate the scattering form factor is the dotted curve in Fig.~\ref{fig_4}~(b). The dashed curve in Fig.~\ref{fig_4}~(d) is normalized by the peak intensity of the solid curve. The region surrounded by dotted vertical lines in Fig.~\ref{fig_4}~(c) corresponds to the horizontal axis of Fig.~\ref{fig_4}~(d). We recognize three peaks of the spectrum in Fig.~\ref{fig_4}~(d). A peak at $\Delta k \sim 0$ is ion feature, and two peaks at $\Delta k \sim \pm \omega_{pe} / c$ are the electron feature. The dotted vertical lines are the peak wavenumbers from Eq.~(\ref{eq_k}), which are consistent with the peaks corresponding to the electron feature. As shown in Fig.~\ref{fig_1}~(a), the two peaks of ion feature are generally much larger than those of electron feature, however, we focus on the electron feature in this study and our computation time is not long enough to properly describe the ion acoustic waves. 

Figure~\ref{fig_5} shows the same plots as Fig.~\ref{fig_4} except with $v_{d2} = 0.01 c$. Since $v_{d2} < v_{te1}$, the plasmas are rather stable as in Fig.~\ref{fig_5}~(a) and do not change the electron distribution function as in Fig.~\ref{fig_5}~(b). As it is stable, Fig.~\ref{fig_5}~(c) is similar to Fig.~\ref{fig_4}~(c). In Fig.~\ref{fig_5}~(d), we can recognize the damping of the right peak due to the smaller derivative of the electron distribution function at the corresponding velocity in Fig.~\ref{fig_5}~(d). The two-plasma theory well represents the numerical spectrum in Fig.~\ref{fig_5}~(d).

Figure~\ref{fig_6} shows the same plots as Fig.~\ref{fig_4} except with $v_{d2} = 0.02 c$, where the plasmas are unstable and the two-stream instability can take place. The two-stream instability grows at $t \sim 20 / \omega_{pe}$, and then saturates. In Fig.~\ref{fig_6}~(b), the positive slope at $v \sim 0.015 c$ at the beginning becomes flatter and forms the plateau at the end of the computation time as a result of the instability. Figure~\ref{fig_6}~(c), which is schematically shown in Fig.~\ref{fig_1}~(b), also indicates the excitations of two-stream instability. The peak intensity of solid curve in Fig.~\ref{fig_6}~(d) is much larger than those in Fig.~\ref{fig_4}~(d) and Fig.~\ref{fig_5}~(d). Although the theoretical spectrum shows similar tendency to the simulated spectrum, where the left peaks have similar value to that in Fig.~\ref{fig_4}~(d) and the right peaks are enhanced, the solid curve is much larger than the dashed one at the right peak. In this figure, we additionally plot the simulated spectrum as the dotted curve using the electron density fluctuations in the latter half of the computation time, which is after the saturation of the two-stream instability. Since the distribution function is flat in the latter half of the computation time, the peak intensity after the saturation is higher than that using all the computation time. Please note that the intensities of the ion feature in Figs.~\ref{fig_4}~(d), \ref{fig_5}~(d), and \ref{fig_6}~(d) are similar value ($\sim 7\times 10^{-6}$), i.e., the range of the vertical axis in Fig.~\ref{fig_6}~(d) is 10 times larger than those in Figs.~\ref{fig_4}~(d) and \ref{fig_5}~(d). 

Comparing the simulated spectra of $v_{d2}=0$ and $0.01c$ in Fig.~\ref{fig_4}~(d) and Fig.~\ref{fig_5}~(d), the peaks at $\Delta k \sim -\omega_{pe}/c$ are similar, while the peak at $\Delta k \sim \omega_{pe}/c$ in the solid curve is larger than that in the dashed curve. This tendency is consistent with the theoretical spectra when $v_{d2} < v_{te1}$. It is considered that the electron distribution functions overlap and Landau damping becomes strong at the peak of $\Delta k \sim \omega_{pe}/c$. The electron density fluctuation in Fig.~\ref{fig_5}~(a) and electron distribution function in Fig.~\ref{fig_5}~(b) does not change over time. Thus, the theoretical function is appropriate to express the shape of spectrum when $v_{d2} < v_{te1}$. The electron density fluctuation in Fig.~\ref{fig_6}~(a) shows the excitations of two-stream instability when $t \gtrsim 20 / \omega_{pe}$. As the peak phase velocity is $\sim 0.017 c$ from the dispersion relations, the drift velocity of $0.02 c$ is reasonable to enhance Langmuir wave via two-stream instability. Thus, it is feasible to diagnose two-stream instability directly via CTS measurements. At the end of the computation time, the electron distribution function in Fig.~\ref{fig_6}~(b) shows the plateau. This is quasi-equilibrium and the slopes is $\sim 0$ in the electron distribution function. Since the derivative of the electron distribution function at the peak phase velocity is $\sim 0$, the right peak of theoretical spectrum in Fig.~\ref{fig_6}~(d) is enhanced. In quantitative sense, the distribution function approximated by two Maxwellian distributions can deviate from the condition of $\partial f/\partial v\sim 0$. Therefore, the numerical result shows larger peak than that of theory in Fig.~\ref{fig_6}~(d). In such quasi-equilibrium plasmas, the scattering form factor in Eq.~(\ref{eq_form}) with two plasmas qualitatively explains the numerical spectra.

\section{Experiment}
\label{sec_experiment}

\begin{figure} 
    \centering
    \includegraphics[clip,width=\hsize]{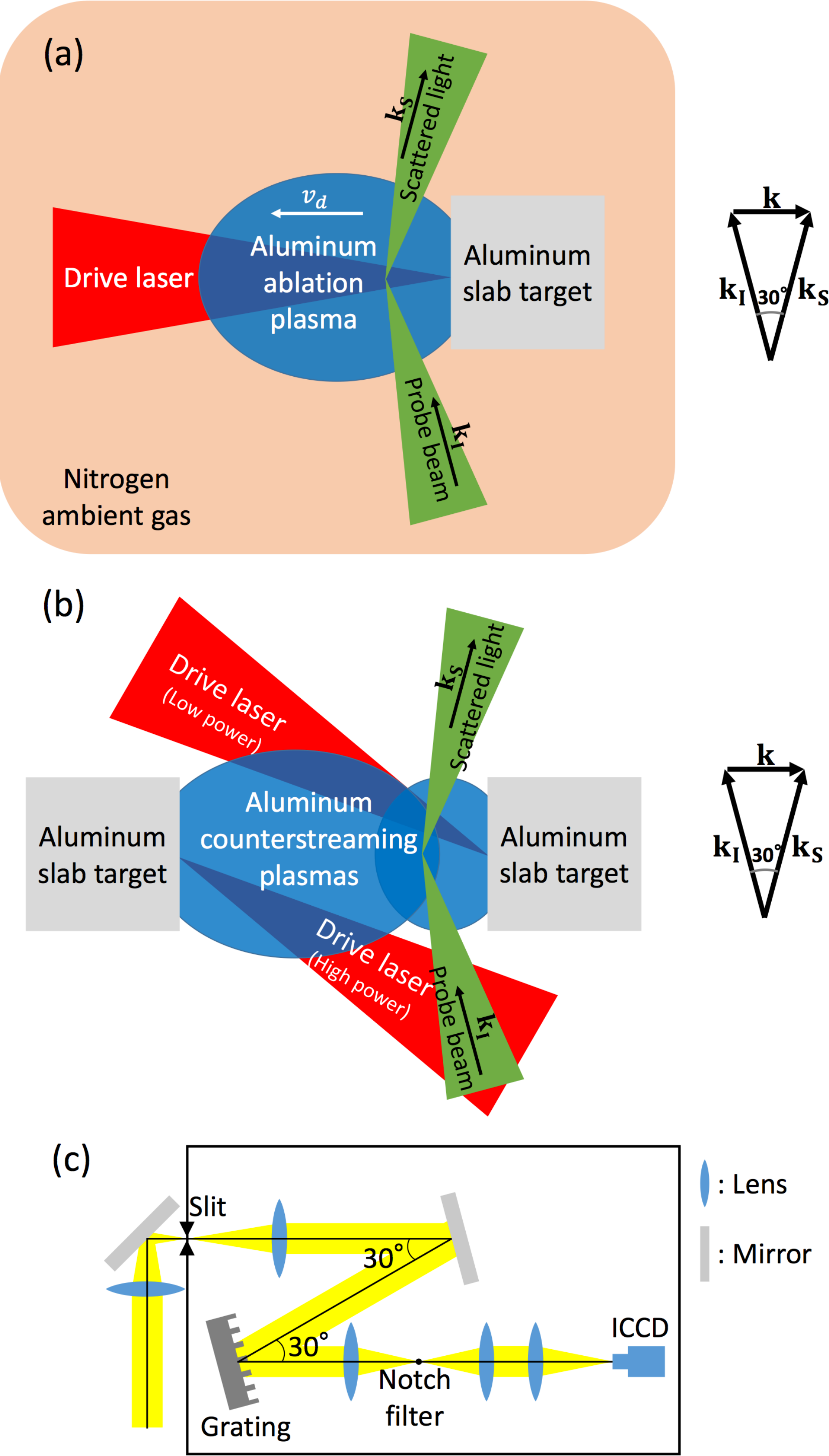} 
    \caption{(a)~Schematic view of the experiment with a single target. (b)~Schematic view of the experiment with a double target. (c)~The design of the CTS spectrometer \cite{bolouki2019hedp}.}
    \label{fig_7}
\end{figure}

In order to verify and develop CTS in the non-equilibrium plasmas, we have designed and conducted the experiments with 100~TW laser facility at National Central University (NCU 100~TW), which is relatively small but high repetition laser with flexible beam lines \cite{hung2014apb,kuramitsu2015hedp}. Figures~\ref{fig_7}~(a) and (b) show the schematic top view of our experimental system. Figure~\ref{fig_7}~(a) shows the setup with a single target. We use the drive laser with the wavelength of 810~nm, the energy of 3.3~J, the uncompressed pulse duration of 150~ps, and the intensity of $\sim 1\times10^{15}$ $\mathrm{W/cm^{2}}$. Without pulse compression, the intensity is a similar level to that of GXII, and we expect to generate a plasma with the similar temperature to GXII. The drive laser is focused on an aluminum slab target. We define the position of 1~mm upstream of the drive laser from the focal position as the reference point, the target chamber center (TCC). The target is irradiated with the drive laser and a plasma is created from the target as shown in Fig.~\ref{fig_7}~(a). The target chamber is filled with nitrogen gas to generate two-stream state where aluminum and nitrogen plasmas coexist. The radiation from the interaction between the aluminum target and the drive laser ionizes the nitrogen gas to produce an ambient plasma (Fig.~\ref{fig_7}~(a)). We move the target by the motional stage every 20~shots to irradiate a new planer surface with the drive laser. 

In order to measure the local plasma quantities, we use the CTS measurement system \cite{bolouki2019hedp}. We use an independent probe beam for CTS in the direction of 105~degrees from the direction parallel to the drive laser propagation, as shown on the right side in Fig.~\ref{fig_7}~(a). We use Nd:YAG laser as the probe beam with the wavelength of 532~nm, the energy of 50~mJ, the pulse duration of 5~ns, and the focal spot size of 100~$\mathrm{\mu}$m. The probe beam is focused at TCC and is scattered by the plasmas. Due to the limitation of the target chamber, we observe the scattered light in the direction of 30~degrees from the direction the probe beam propagates. The CTS system measures the local parameters of plasmas in the direction parallel to the drive laser as shown on the right side in Fig.~\ref{fig_7}~(a). The scattered light is detected with the spectrometer of CTS in Fig.~\ref{fig_7}~(c) \cite{bolouki2019hedp}. The image of probe beam is transferred through the slit to the detector of ICCD keeping the spatial information. We use a reflective holographic grating (1200 grooves/mm) to spectrally resolve the electron feature. We put a notch filter after the grating to remove the stray light at the same wavelength as the probe beam. The spectrum of the scattered light is obtained and accumulated 20 times using ICCD camera with the gate width of 2~ns. The spatial and temporal resolutions are determined by the focal spot size of the probe beam and the gate width of ICCD camera, respectively. 

Figure~\ref{fig_7}~(b) shows the setup with a double target. The setup of the right target and CTS is same as that in Fig.~\ref{fig_7} (a). We put another aluminum slab target 5~mm away from the existing target surface to generate counterstreaming plasmas. Even though the laser energy is not large enough to excite shock, we are still able to study the counterstreaming plasmas to develop the diagnostics with NCU 100~TW. We divide the drive laser beam into two beams with the beam splitter. The two beams are focused on the left and right targets, respectively. The targets are irradiated with the drive laser beams at the angle of 30 degrees from the target normal as shown in Fig.~\ref{fig_7}~(b). Two ablation plasmas created by both beams have different velocities and propagate in the opposite directions, hence the two plasmas contact at a point. The contacting point is placed on TCC and we diagnose the non-equilibrium counterstreaming plasmas via CTS. 

\begin{figure*}
    \centering
    \includegraphics[clip,width=\hsize]{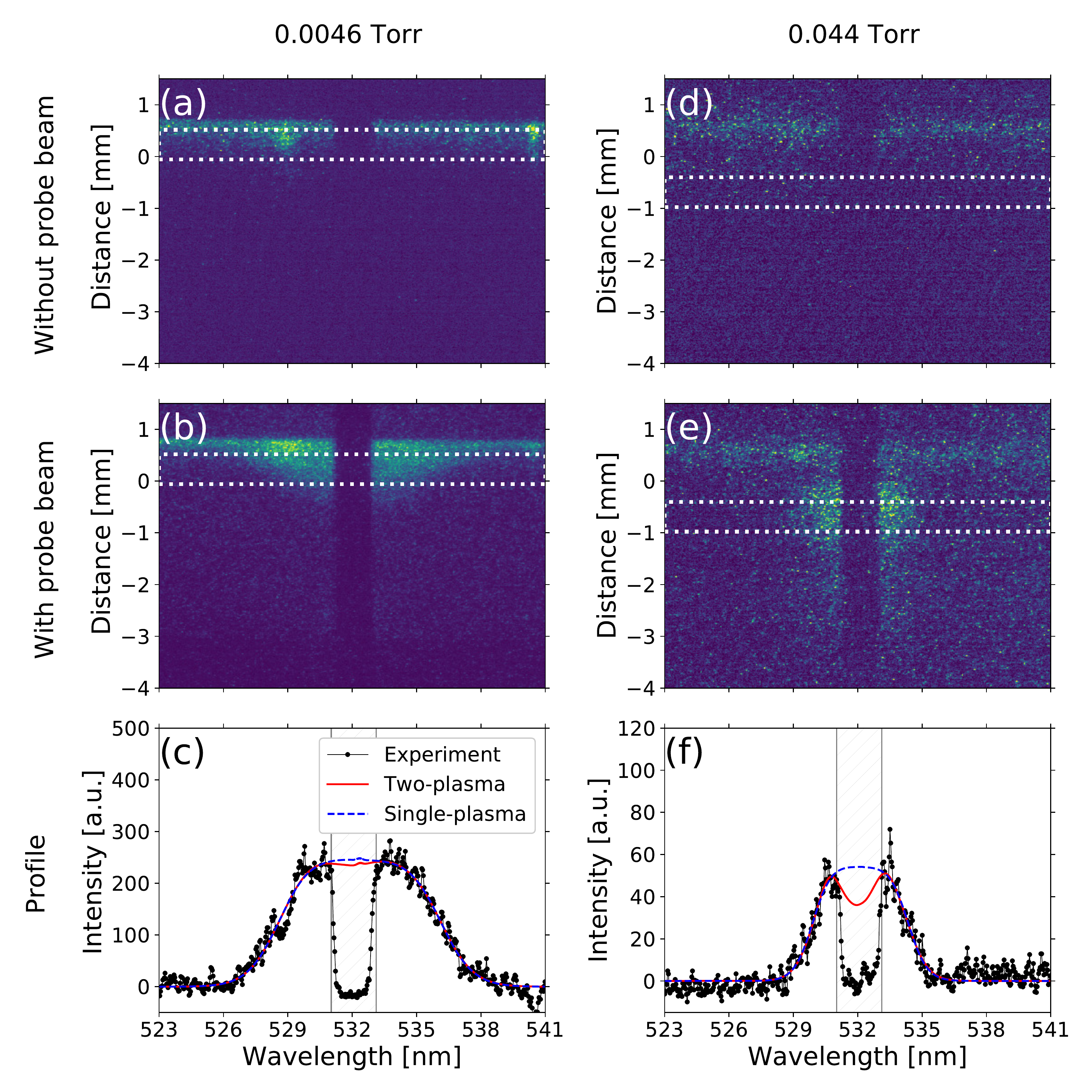}
    \caption{Experimental results. (a), (d) and (b), (e) are the image of CTS spectrometer without and with probe beam. (c) and (f) are the profiles of (b) and (e), respectively. (a)-(c) and (d)-(f) are the results 30~ns and 70~ns after the drive laser irradiation, at the pressures of 0.0046~Torr and 0.044~Torr, respectively.}
    \label{fig_8}
\end{figure*}

\begin{table*}
    \centering
    \caption{Fitting results.}
    \begin{tabular}{c|cccccccccc} \hline
        $P$ [Torr]& &$T_{e1}$ [eV]&$T_{e2}$ [eV]&$n_{e1}$ [$10^{16} \mathrm{cm^{-3}}$]&$n_{e2}$ [$10^{15}\mathrm{cm^{-3}}$]&$Z_{1}$&$Z_{2}$&$v_{d1}$ [km/s]&$\alpha_{1}$&$\alpha_{2}$\\ \hline
        \multirow{2}{*}{0.0046} &$S_{1}$  &$30.1\pm0.9$ &              &$2.66\pm0.13$ &       &$5.40$ &       &$383\pm23$&$0.654$ & \\
                                &$S_{1+2}$&$30.4\pm1.1$ &$6.76\pm8.68$ &$2.69\pm0.25$ &$0.85$ &$5.42$ &$3.02$ &$391\pm23$&$0.655$ &$0.246$ \\ \hline
        \multirow{2}{*}{0.044}  &$S_{1}$  &$10.3\pm0.7$ &              &$0.91\pm0.14$ &       &$3.13$ &       &$134\pm23$&$0.654$ & \\
                                &$S_{1+2}$&$10.9\pm1.0$ &$2.63\pm1.36$ &$1.03\pm0.26$ &$3.14$ &$3.20$ &$1.11$ &$161\pm28$&$0.676$ &$0.760$ \\
        \hline
    \end{tabular}
    \label{exp_param}
\end{table*} 

Here we show the preliminary experimental results with a single target. Figures~\ref{fig_8}~(a)-(c) and (d)-(f) show the results of CTS with 0.0046~Torr and 0.044~Torr of the ambient gas at the timing of 30~ns and 70~ns, respectively. There is the result at the same timing, however, the signal is the CTS signal is very weak and noisy. Thus, the results with different timing are shown here. Figures~\ref{fig_8}~(a), (d) and (b), (e) are the images of CTS without and with the probe beam, respectively. The vertical and horizontal axes of each image represent the distance along the probe beam and wavelength, respectively. The probe beam propagates from upper side to lower side of the vertical axis. The position where the distance equals to zero is TCC. There is no signal from 531~nm to 533.2~nm due to the notch filter in Fig.~\ref{fig_8}. It is possible to observe only emission from plasma such as bremsstrahlung without the probe beam in Figs.~\ref{fig_8}~(a), (d) and both emission and CTS with the probe beam in Figs.~\ref{fig_8}~(b), (e). The CTS signals are found in Figs.~\ref{fig_8}~(b) and (e), and they are not found in Figs.~\ref{fig_8}~(a) and (d). Figures~\ref{fig_8}~(c) and (f) show the CTS profiles at the distance of $0.23$ and $-0.69$ mm, respectively. To reduce the emission, we electronically subtract the profiles in Figs.~\ref{fig_8}~(a) and (d) from those in Figs.~\ref{fig_8}~(b) and (e), respectively. We averaged the signals over vertical 50~pixels (575~$\mathrm{\mu m}$) to reduce the noise and enhance the signal to noise ratio. The regions of images to make the profiles are shown in the dotted rectangles in Figs.~\ref{fig_8}~(a), (b), (d), and (e). The curves with markers are the experimental data. The solid and dashed curves represent the fitting result with two-stream theoretical function in Eqs.~(\ref{eq_form}), (\ref{eq_fe}), and (\ref{eq_chie}) (two-plasma fitting), and that with the conventional analysis method (single-plasma fitting), respectively. The notch filter is shaded in Figs.~\ref{fig_8}~(c) and (f). Although the width of notch filter in Figs.~\ref{fig_8} (c) and (f) looks slightly larger than that shown in the images of Figs.~\ref{fig_8} (a), (b), (d), and (e), we define the width in Figs.~\ref{fig_8} (c) and (f) using the image without CTS signals in order to prevent the effect of notch filter in the fitting. We choose the shaded width where the broad signal begins to decrease. We fit the result by the least-squared method in the region without the notch filter to estimate the parameters and the errors. The fitting variables are $T_{e1}$, $T_{e2}$, $n_{e1}$, and $v_{d1}$ in the two-plasma fitting, while $T_{e1}$, $n_{e1}$, and $v_{d1}$ in the single-plasma fitting. The $j=1$ and $j=2$ plasma species are aluminum and nitrogen, respectively. The ion valences ($Z_{1}$ and $Z_{2}$) are calculated with FLYCHK \cite{flychk}. The electron density from nitrogen is expressed as $n_{e2} = Z_{2} n_{i2}$, where $n_{i2}$ is the nitrogen density. The nitrogen density at the pressures of 0.0046~Torr and 0.044~Torr are $2.8 \times 10^{14} ~ \mathrm{cm^{-3}}$ and $2.8 \times 10^{15} ~ \mathrm{cm^{-3}}$, respectively. The aluminum plasma is assumed to be used in the single-plasma fitting. We assume the ambient nitrogen plasma at rest in laboratory, $v_{d2} = 0$, in the two-plasma fitting. 

Table~\ref{exp_param} shows the parameters and errors obtained by the fittings. The parameters are averaged over 575~$\mathrm{\mu m}$ in the direction of probe beam since we averaged the signal over 50~pixels. Although the aluminum parameters are similar with both fitting methods, the changes in parameters between the fitting methods are larger in 0.044~Torr case. In Figs.~\ref{fig_8}~(c) and (f), the solid curves are in agreement with the dashed ones at the position away from the notch filter, however, the solid curve is a little different from the dashed ones near the notch filter in Fig.~\ref{fig_8}~(f). From the fitting results shown in Table~\ref{exp_param}, the scattering $\alpha_{j}$ are less than 1 at both pressures due to the low electron density. 

The parameters of electron density and temperature shown in Table~\ref{exp_param} obtained from NCU 100~TW experiment are as the same order as that obtained with high power lasers which is typically used in laboratory astrophysics \cite{bolouki2019hedp}. The result shows that it is possible to conduct experiments of laboratory astrophysics using tabletop lasers. We constructed the CTS electron feature spectrometer and obtained CTS spectra at NCU 100~TW laser facility. Our plasma is not completely non-collective. The $\alpha < 1$ but close to 1, so there are still collective feature as shown in Fig.~\ref{fig_3} in the Sec. \ref{sec_theory}; superposition of two "rather non-collective" plasmas result in two peaks in scattered spectra with two-plasma theory. In the experiment, we observed the region where the drift velocity was quite smaller than the thermal velocity and there is little difference between two fitting methods. The electron distribution function became almost symmetric about $v=0$ and there were little difference of intensity among peaks. However, there are some differences in the spectra between the two-plasma and single-plasma fittings in Fig.~\ref{fig_8}~(f). Although the results are limited, the intensity in Fig.~\ref{fig_8}~(f) begins to decrease toward the central wavelength outside the notch filter unlike that in Fig.~\ref{fig_8}~(c). It is considered that the electron temperatures shown in Table~\ref{exp_param} are different for different plasma species, and that the electron distribution function has more low energy component than Maxwellian. As the electron density from aluminum and nitrogen plasma have similar values, especially at the pressure of 0.044~Torr, the influence of nitrogen plasma on the spectrum becomes large and the difference appears in the fitted curves. However, since the results are limited, further experiments are required to verify this. 

We have generated plasmas with similar temperature and density to that produced with large lasers such as GXII, but with smaller volume due to the much smaller energy of the tabletop laser. We are planning to observe the plasma dynamics by the uncompressed laser pulse and to compare the dynamics to that in high-power laser pulse in the future. We have also obtained electron features of CTS from the plasmas. When the ambient plasma density is higher, the two-plasma fitting spectrum shows finite difference from the single-plasma fitting. This may be a detection of the abundance of lower energy electrons in distribution function. Although the obtained plasma parameters are still mostly non-collective, we will increase the ambient gas pressure or use counterstreaming plasmas to obtain higher density plasma in the future. Moreover, we will change the scattering angle to observe two-stream instability. 

\section{Summary}
\label{sec_summary}

In summary, we investigate the electron feature of CTS in two-plasma states theoretically, numerically, and experimentally. The theoretical analysis in the presence of two-stream plasmas show qualitative modifications of the CTS spectra due to the derivative of electron distribution functions at the resonant velocities. When $v_d<v_{te}$, the peak in $\Delta k > 0$ will be damped due to the Landau damping. On the other hand, when $v_{d}>v_{te}$, Landau resonance is not well expressed. Therefore, we numerically calculate the spatio-temporal evolutions of two-stream plasmas with PIC simulations, and then, numerically solve the wave equation for the scattered waves using the density fluctuations from PIC simulations as the source of the wave equation. From the numerical simulations, the results when $v_d<v_{te}$ are consistent with those in the analytical investigations. When $v_d>v_{te}$, it is considered that the two-stream instability also affected CTS spectra. After the saturation of two-stream instability, the electron distribution function forms a plateau at the positive velocity wing, which is a quasi-equilibrium state. In this state, the derivative of electron distribution function $\sim 0$ at the peak phase velocity and one of the peaks of theoretical spectrum is enhanced, which is qualitatively consistent with that of the simulated spectrum. When $v_{d1} = v_{d2}$ and $T_{e1} \neq T_{e2}$, the low and high energy components of the electron distribution functions affect the scattering spectra. In order to verify the analytic and numerical investigations of CTS in the presence of two-stream plasmas and instability, we have been developing a down scaled experimental system for laboratory astrophysics with a relatively small laser facility, the 100~TW laser facility at National Central University with an uncompressed laser pulse. Once we fully understand the non-equilibrium CTS, we can use the CTS as diagnostics not only for the non-equilibrium plasmas but also for the instabilities. 

\begin{acknowledgments}

This work is supported by the Ministry of Science and Technology of Taiwan under grant no.~103-2112-M-008-001-MY2, 104-2112-M-008-013-MY3, and 105-2112-M-008-003-MY3, and JSPS KAKENHI Grant number JP15H02154, JP17H06202, and JP18H01232. Partially supported by JSPS Core-to-Core Program B. Asia-Africa Science Platform Grant No. JPJSCCB20190003 

\end{acknowledgments}

\section*{Data availability}
The data that support the findings of this study are available from the corresponding author upon reasonable request. 

\appendix

\begin{figure*}
    \centering
    \includegraphics[clip,width=\hsize]{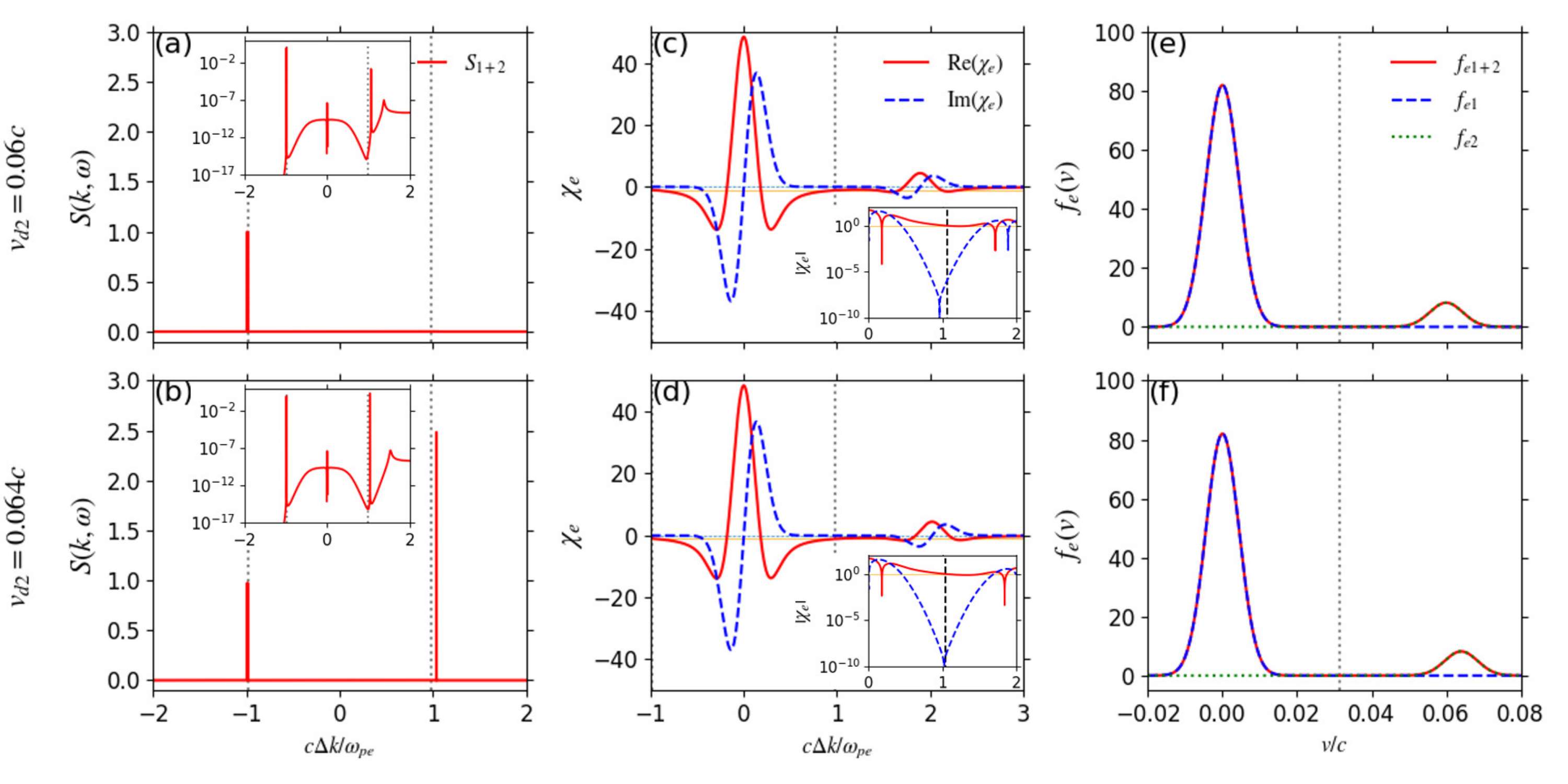}
    \caption{(a), (b) Theoretical CTS spectra from two-stream plasmas with the scattering angle of 30~degrees. (c), (d) Electron susceptibilities. (e), (f) Electron distribution functions. (a), (c), (e) $v_{d2}=0.06c$, (b), (d), (f) $v_{d2}=0.064c$.}
    \label{fig_9}
\end{figure*}

\section{Scattering form factor in non-equilibrium plasmas}
\label{sec_appendix}

Figures~\ref{fig_9}~(a) and (b) show more collective spectra due to the smaller scattering angle (30~degrees). The inserted figures in Figs.~\ref{fig_9}~(a) and (b) show the plot where the y-axis is logarithmic. As shown in inserted figure in Fig.~\ref{fig_9}~(a), there is actually right peak but strongly attenuated. Although the right peaks in Figs.~\ref{fig_9}~(a) and (b) are different, the distribution functions in Figs.~\ref{fig_9}~(e) and (f) are similar. Since both distribution functions in Figs.~\ref{fig_9}~(e) and (f) are unstable, the instability grows similarly in both cases. Figures~\ref{fig_9} (c) and (d) show the electron susceptibilities. We plot absolute value of the electron susceptibilities in the lower right. The vertical dashed lines in the lower right figures show the wavenumber where $\mathrm{Re} (\chi_{e})=-1$ and they correspond to the peak wavenumber. Although the imaginary parts of electron susceptibility in Figs.~\ref{fig_9}~(c) and (d) seem to have similar values at the right peak, the value in Fig.~\ref{fig_9}~(c) is $\sim 10^{3}$ times larger than that in Fig.~\ref{fig_9}~(d) as shown in the lower right figures. Due to the superposition of distribution functions, the real and imaginary parts of $\chi_e$ are close to $-1$ and $0$ at the right peak in Fig.~\ref{fig_9} (d), respectively. In such case, since the denominator of $|1-(\chi_e/\epsilon)|$ approaches to $0$, the peak intensity is mathematically enhanced. As $\chi_e$ is closer to the resonant condition with the relative drift velocity of $0.065 c$, the peak intensity is $\sim 20$ times higher than that in Fig.~\ref{fig_9}~(b) (not shown). Thus, the peak intensity is sensitive to $\chi_e$ and rapidly changes. While the negative slope of the distribution function relevant to the Landau damping is included in the theory, the instability is not. The right peak enhancement in Fig.~\ref{fig_9} is nothing to do with the instability but simply mathematically determined. 

\bibliography{netse}

\end{document}